%% file: main.tex
\documentclass[a4paper, amsfonts, amssymb, amsmath, reprint, showkeys, nofootinbib, twoside, floatfix, superscriptaddress]{revtex4-1}

\input{preamble}

\begin{document}
\title{Estimates for Disk and Ejecta Masses Produced in Compact Binary Mergers}

\author{Christian J. Kr\"uger}
    \email{christian.krueger@tat.uni-tuebingen.de}
    \affiliation{Department of Physics, University of New Hampshire, 9 Library Way, Durham, NH 03824, USA}
    \affiliation{Theoretical Astrophysics, IAAT, University of T\"ubingen, 72076 T\"ubingen, Germany}
\author{Francois Foucart}
    \email{francois.foucart@unh.edu}
    \affiliation{Department of Physics, University of New Hampshire, 9 Library Way, Durham, NH 03824, USA}

\date{\today} 

\begin{abstract}
There is irresistible observational evidence that binary systems of compact objects with at least one neutron star are  progenitors of short gamma-ray bursts, as well as a production site for r-process elements, at least when some matter is ejected by the merger and an accretion disk is formed. The recent observations of gravitational waves in conjunction with electromagnetic counterparts fuel the need for models predicting the outcome of a given merger and the properties of the associated matter outflows as a function of the initial parameters of the binary. In this manuscript, we provide updated fitting formulae that estimate the disk mass for double neutron star binaries and ejecta masses for black hole-neutron star and double neutron star binaries, fitted to the results of numerical simulations. Our proposed fitting formulae improve on existing models by aiming for analytical simplicity, by covering a larger region of parameter space, and by accounting for regions of parameter space not covered by numerical simulations but with physically manifest merger outcomes.
\end{abstract}


\maketitle

\section{Introduction} \label{sec:introduction}

The new era of gravitational wave (GW) astronomy has been heralded by exciting observations of binary coalescences of compact objects. At least two of the events, namely GW170817~\cite{2017PhRvL.119p1101A,2017ApJ...848L..12A,2017ApJ...848L..13A} and GW190425~\cite{2020ApJ...892L...3A}, are likely double neutron star mergers, and the observations of electromagnetic (EM) signals following GW170817 clearly indicates the presence of at least one neutron star in that system. With KAGRA coming online in the near future, extending the currently operating trio of the LIGO and Virgo detectors, and considering planned upgrades for existing detectors, we are expecting not only the detection rate of such GW events to increase, but also to substantially improve the localisation of those events in the sky.\cite{2018LRR....21....3A} This will improve our chances of performing joint EM and GW observations of these events, increasing the scientific return of GW observations.

When a neutron star is torn apart by the tidal forces of its black hole companion or collides with another neutron star, most of its material ends up within the post-merger remnant compact object. However, a small fraction of the neutron star ($\sim 0.01M_\odot-0.3\,M_\odot$) may be dynamically ejected from the system or form an accretion disk around that compact object. It is these debris that fuel EM transients such as kilonovae~\cite{1998ApJ...507L..59L,1976ApJ...210..549L,2010MNRAS.406.2650M} and short gamma-ray bursts (SGRBs) \cite{1986ApJ...308L..43P,1989Natur.340..126E}. The intensity of these EM transients and other observables strongly depends on the amount of matter that is ejected during merger, $M_{\rm dyn}$, bound in an accretion disk, $M_{\rm disk}$, or ejected in the form of post-merger disk outflows. Both $M_{\rm dyn}$ and $M_{\rm disk}$ depend on the properties of the coalescing compact objects~\cite{Foucart:2012nc,Kawaguchi:2016ana,2017CQGra..34j5014D,2018ApJ...869..130R}, while the fraction of the disk mass unbound in disk winds (up to $\sim 0.5M_{\rm disk}$) also strongly depends on the large scale structure of magnetic fields in the post-merger remnant~\cite{Christie:2019lim}.

Fitting formulae in general, and those for disk and ejecta masses in particular, are valuable tools with many potential applications. They provide predictions for quantities that would otherwise only be accessible via computationally expensive numerical simulations performed in full General Relativity. Owing to the variety of parameters for such simulations (masses and mass ratio of the two compact objects, spin and spin alignment, equation of state, etc.), simulations cover only a subset of the possible parameters and make interpolations and extrapolations to yet unexplored regions of the parameter space necessary. Formulae estimating disk or ejecta masses are already used to assess the usefulness of triggering EM follow-up searches to GW events~\cite{2012PhRvD..86l4007F,2014ApJ...791L...7P,2019arXiv191100116C}. Alternatively, they can be ``inverted'' and used (in conjunction with other observations) to constrain the parameters of a binary system after the observation of EM signals (such as SGRBs or kilonovae)~\cite{2018ApJ...852L..29R,Hinderer:2018pei,2019MNRAS.489L..91C,Barbieri:2019bdq,Coughlin:2019zqi,2020ApJ...890..131A,2019ApJ...877...94A}. 

Here, we propose fitting formulae for the following three particular cases: disk mass from binary neutron star (BNS) mergers, and dynamical ejecta masses for BNS and black hole-neutron star (BHNS) mergers. We very recently updated our model for the disk mass resulting from BHNS mergers~\cite{2018PhRvD..98h1501F}, and do not attempt to improve it further here. For the other three cases, we review existing fitting formulae and propose improvements that reduce their analytical complexity, and/or increase their range of validity by calibrating them to a broader dataset. This is done in part by getting rid of some terms in the fitting formulae that were originally derived from physical considerations applying to the disruption of BHNS binaries, but do not necessarily apply to BNS systems; by taking into account the desired behavior of these formulae for very compact stars, even in the absence of numerical simulations in that region of parameter space; and by taking advantage of some newly released numerical simulations.

Throughout this paper we work in units in which $c = G = M_\odot = 1$.

\section{Disk Mass for BNS Binaries} \label{sec:nsns_diskmass}

\subsection{Existing fitting formulae}

Radice \etal\cite{2018ApJ...869..130R} (henceforth REA) performed a comprehensive survey on the mass ejections and the associated electromagnetic transients from binary neutron star mergers. Their survey is based on 35 numerical relativity (NR) simulations,
employing four different realistic equations of state fulfilling current astrophysical constraints, and covering a large variety of neutron star masses for binary neutron stars; within their set of binaries the mass ratio, $q = M_1/M_2$, is confined to $0.86 \le q \le 1$. We show relevant data of those simulations in Table~\ref{tab:bns_diskmass_data} in the Appendix.

For these simulations, they find that the remnant disk mass, $M_{\rm disk}$, is to good approximation a function of the binary's effective dimensionless tidal deformability, $\tilde{\Lambda}$~\cite{2008PhRvD..77b1502F,2014PhRvL.112j1101F}, and can be modeled as
\begin{align}
    \frac{M^\text{REA}_\text{disk}}{M_\odot}
        = \max \left\{
        10^{-3},
        \alpha + \beta \tanh \left(
            \frac{\tilde{\Lambda}-\gamma}{\delta}
            \right)
        \right\},
\end{align}
with $\alpha = 0.084$, $\beta = 0.127$, $\gamma = 567.1$, and $\delta = 405.14$. The formula predicts that for a binary with tidal deformability $\tilde{\Lambda} \lesssim 250$ hardly any disk forms. With increasing tidal deformability, more material assembles to form a disk and for $\tilde{\Lambda} \gtrsim 750$ the disk mass levels off at $\approx 0.2\,M_\odot$.

The same set of binary simulations as referred to above was used by Coughlin \etal\cite{2019MNRAS.489L..91C} (henceforth CEA), who developed an alternative formula based on the idea that the lifetime of the remnant prior to collapse to a black hole is mostly governed by $M_\text{tot}/M_\text{thr}$, where $M_\text{tot}$ is the total mass of the binary and $M_\text{thr}$ is the threshold mass, above which the merger results in prompt collapse to a black hole, as defined in \cite{2013PhRvL.111m1101B}. They find that the model
\begin{align}
    & \log_{10}\left(
        \frac{M^\text{CEA}_\text{disk}}{M_\odot}
    \right) = \nonumber\\
    & \quad \max\left\{
            -3,\,
            a \left(
                1 + b \tanh\left[
                    \frac{c-M_\text{tot}/M_\text{thr}}{d}
                \right]
            \right)
        \right\}
\end{align}
provides an accurate description for the data from the NR simulations, with the coefficients $a = -31.335$, $b  = -0.9760$, $c = 1.0474$, and $d = 0.05957$. The fitting formula suggests that the merger of a binary with total mass of $M_{\rm tot} \ge 0.95 M_{\rm thr}$ will not result in a significant accretion disk. There are noticeable differences in the prediction of these two formulae, yet they are not as dissimilar as they might initially appear: as a rule of thumb, equations of state with larger $M_{\rm thr}$ also lead to larger $\tilde \Lambda$.

After the development of these analytical predictions, Kiuchi \etal\cite{Kiuchi:2019lls} reported disk masses for 22 NR simulations using polytropic equations of state. Importantly, these simulations include BNS mergers with asymmetric mass ratios (they report results for $q=0.775$ and $q=1$), with outcomes that are not always well captured by existing fitting formulae developed for nearly equal mass binaries. We find that both above mentioned formulae work well for the dataset compiled by Radice \etal, i.e. for the parameter range that they were intended to cover. However, outside these ranges, the estimates for the disk mass becomes less precise. This led us to investigate if we can find a fitting formula that works well for the combined set of NR simulations. When deriving new fitting formulae, we will thus combine the data from Radice \etal and Kiuchi \etal\cite{2018ApJ...869..130R,Kiuchi:2019lls} which we list in Table~\ref{tab:bns_diskmass_data} in the Appendix.\footnote{Although for~\cite{2018ApJ...869..130R} we consider only simulations performed at the reference resolution $h = 185\unit{m}$ and without neutrino heating.} The neutron star spin in all considered simulations is zero.

\subsection{Proposed fitting formula}

A generic issue with finding fitting formulae for such data is the relatively large error bars that are attached to many of the  quantities derived from NR simulations. A fraction of these errors are, obviously, of numerical nature as the resolution of the simulations is limited and modeling microphysics adds to the computational expense, often at the cost of accuracy. On the other hand, some quantities, like the remnant disk mass in the case of binary neutron stars that we are interested in, suffer from the lack of an unambiguous definition: in the immediate aftermath of a violent, disruptive neutron star merger in which matter is strongly redistributed, the question arises on how to distinguish between the ``remnant object'' and its surrounding ``accretion disk''.  Finally, important physical processes are still ignored, approximated, or not resolved in simulations, potentially affecting the properties of the post-merger remnant. \footnote{E.g. neutrino heating, magnetic fields and the associated magneto-rotational instability.} These contributing factors lead us to assuming an error of
\begin{equation}
    \Delta M_\text{disk}
         = 0.5 M_\text{disk} + 5 \cdot 10^{-4} M_\odot
    \label{eq:errorbars}
\end{equation}
when fitting the numerical data. In practice, these errors determine the relative weight of various numerical simulations in our fitting procedure.

We considered a few alternative forms of the fitting formulae that may perform better for asymmetric binaries. In particular, it seemed that in the high mass ratio regimes, our formula for BNS systems may become similar to the well working fitting formula for the remnant mass of a BHNS merger~\cite{2018PhRvD..98h1501F}, as massive neutron stars are extremely compact. To our surprise, however, a rather simple fitting formula allows us to predict the disk mass for our reference numerical simulations to good accuracy:
\begin{equation}
    M_\text{disk}
        = M_1 \max\left\{5 \times 10^{-4}, \left(a C_1 + c\right)^d  \right\},
    \label{eq:nsns_disk}
\end{equation}
where $C_1 = GM_1/(R_1c^2)$ is the compactness of the lighter of the two neutron stars, $M_1$ its gravitational mass, and $R_1$ its radius; our calibration dataset (cf. Table~\ref{tab:bns_diskmass_data}) covers a range of $C_1 \in \left[0.135,\,0.219\right]$; the range of mass ratios spans $q \in \left[0.775,\,1\right]$. A least squares fit using \eqref{eq:nsns_disk} yields the coefficients $a = -8.1324$, $c = 1.4820$, and $d = 1.7784$.

Extending this formula by adding other terms does not meaningfully improve the quality of the fit. Merely three binary systems out of the 57 NR simulations show significant deviations from our fitting formula; however, those three systems tend to be poorly fitted by all existing analytical formulae. We will pay special attention to them when discussing the quality of our proposed formula in the following Subsection~\ref{ssec:quality}.

Why should such a simple formula work? In the limit of high mass ratios, the disk forms from the tidal tail created by tides in the lower mass neutron star. Our formula matches, within the expected errors, results obtained for BHNS systems at mass ratios $Q\sim 1.5-2$, if one replaces the more massive neutron star by a non-spinning black hole. We should, however, expect some dependence on the mass ratio of the system (as for BHNS binaries) and, for more symmetric binaries, in the properties of the second neutron star. The fact that more advanced formulae, that borrow from the ideas of REA and CEA for symmetric binaries or include some dependence on the properties of the more massive neutron stars, do not, in our experience, provide better fits is most likely a sign of the current limitations of our sparse set of numerical results, and possibly of the impact of significant numerical/modeling uncertainties.

Our model predicts that for $C_1 > 0.182$ no accretion disk is formed. This can be understood in the way that more compact neutron stars do not form significant tidal tails. We note that our fitting formula agrees with REA and CEA in that equal mass binaries with 
small $\tilde \Lambda$ or large $M_{\rm tot}$, which also have large $C_1$, do not form massive remnant disks. Yet it also allows for the formation of an accretion disk when a lower mass, less compact neutron star merges with a massive companion, as seems to happen in higher mass ratio simulations performed by Kiuchi \etal\cite{Kiuchi:2019lls}. Our formula clearly runs into trouble when considering very low compactness: in the limit of $C_1 = 0$, a disk mass of $\approx 2.0 M_1$ is predicted, which is clearly incorrect (nearly all material from both neutron stars would be accumulated in the accretion disk). This, however, poses no serious problem as astrophysical neutron stars should have compactnesses of $C \gtrsim 0.12$. We will be satisfied as long as our fitting formula produces accurate values for physically realistic compactnesses.

\subsection{Quality of Proposed Fitting Formula}
\label{ssec:quality}

\begin{figure}[htbp]
    \centering
    \includegraphics[width=0.48\textwidth]{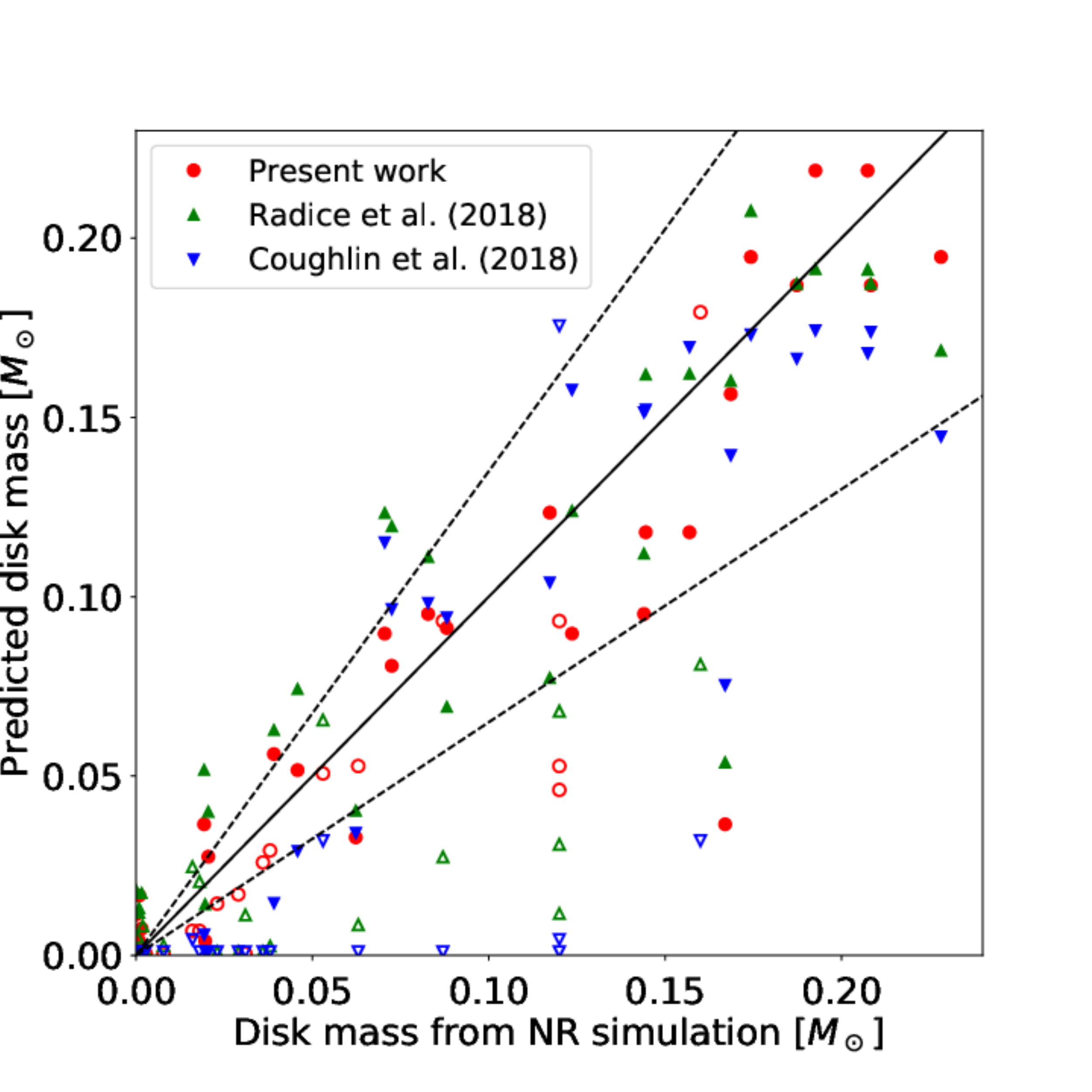}
    \caption{Predicted disk mass vs. disk mass from NR simulations for the two pre-existing formulae (green, upward triangles for REA and blue, downward triangles for CEA) and the proposed formulae (red dots). The two outer, dashed lines depict a 35\,\% deviation from exact prediction. Beside the case of very low disk masses the formulae perform very well (with the exception of the three individually discussed outliers). For added clarity, since both the CEA and REA formulae were not calibrated using the dataset from Kiuchi \etal, we show predictions of the formulae for those data with unfilled symbols.}
     \label{fig:nr_vs_fit}
\end{figure}

We show the disk mass predicted by our proposed fitting formula against the disk mass from the NR simulations in Fig.~\ref{fig:nr_vs_fit}. For the majority of binaries, the fit reproduces the ``measured'' disk mass to an accuracy of better than 35\%. The accuracy naturally becomes worse when considering binaries that form only a very low mass disk; given our rather large uncertainties, cf. Equation~\eqref{eq:errorbars}, we expect such behaviour from virtually any proposed fitting formula.

\begin{table}[htbp]
  \caption{Measured and predicted (by the presently proposed and the two referenced fitting formulae) disk masses of the three outliers. All masses are given in solar masses.}
  \begin{tabular}{llccc}
    \toprule
    Ref + ID & $M^\text{NR}_\text{disk}$
        & $M^\text{present}_\text{disk}$
        & $M^\text{REA}_\text{disk}$
        & $M^\text{CEA}_\text{disk}$ \\
    \midrule
    \cite{2018ApJ...869..130R} \texttt{DD2\_M150150\_LK}
        & 0.167
        & 0.037
        & 0.054
        & 0.075 \\
    \cite{Kiuchi:2019lls} $\Gamma = 3.252$, $q=0.775$
        & 0.12
        & 0.053
        & 0.012
        & 0.001 \\
    \cite{Kiuchi:2019lls} $\Gamma = 2.640$, $q=1$ 
        & 0.12
        & 0.046
        & 0.068
        & 0.176 \\
    \bottomrule
  \end{tabular}
  \label{tab:outliers}
\end{table}

Fig.~\ref{fig:nr_vs_fit} clearly shows the three already mentioned outliers. Those have disk masses of $M^\text{NR}_\text{disk} = 0.12 M_\odot$,  $0.12 M_\odot$ and $0.17 M_\odot$, whereas the proposed formula yields predictions of $(0.037-0.053)\,M_\odot$, i.e. the prediction is smaller by a factor of $3-4$ (cf. Table~\ref{tab:outliers}). We will now discuss them individually.

\begin{itemize}[nosep,leftmargin=0pt,labelindent=\labelsep,itemindent=*]

\item To better understand the (strongest) outlier \texttt{DD2\_M150150\_LK}, let us consider sequences of equal-mass binaries of increasing neutron star masses from~\cite{2018ApJ...869..130R} (at fixed equation of state). We observe that, for most sequences, the reported $M^\text{NR}_\text{disk}$ decreases monotonically as the mass $M_\text{NS}$ of each neutron star increases, as expected. Simulation \texttt{DD2\_M150150\_LK} is the only one that does not fit this pattern: the relevant NR disk masses for the \texttt{DD2} equation of state are $M^\text{NR}_\text{disk} = (15.69, \,12.36, \,\textbf{16.70}, \,1.96) \cdot 10^{-2} M_\odot$ for $M_{\rm NS}=(1.35,1.4,1.5,1.6)M_\odot$ (we highlighted the outlier in bold). This sequence shows that, for this system, a small change in the parameters of the binary may dramatically change the remnant disk mass. As this is the only sequence of equal-mass binaries for which a non-monotonic $M^\text{NR}_\text{disk}$ is reported, but the total number of such sequences remain quite low, it is difficult at this point to provide a definitive answer as to the cause of the large observed $M^\text{NR}_\text{disk}$. A slightly higher than usual numerical error for one of the simulations could easily be the cause of this feature, but a non-monotonic behavior of binaries close to the threshold for rapid collapse to a black hole cannot be ruled out either.

\item Somewhat similar arguments can be made for the outliers from~\cite{Kiuchi:2019lls}, which considers only binaries with a total mass of $M_\text{tot} = 2.75\,M_\odot$ while changing the mass ratio and the equation of state. In~\cite{Kiuchi:2019lls}, the equation of state is characterized by $M_\text{max}$, the maximum mass of a non-rotating neutron star, and $P_{14.7}$, the pressure at a density $\rho=10^{14.7}\,{\rm g/cm^3}$. For the binary $\Gamma = 3.252$, $q=0.775$, we can consider a sequence of binaries with $M_\text{max} = 2.05\,M_\odot$, mass ratio $q=0.775$, and increasing $\log P_{14.7}$. Along this sequence the disk mass increases, but with rapid changes in the disk mass that are not resolved given the sparseness of the available numerical dataset: $M^\text{NR}_\text{disk} = (2.9, \,3.8, \,\textbf{12.0}, \,12.0, \,18.0) \cdot 10^{-2} M_\odot$. We can see once more that the outlier lies in a region of parameter space where a small change of input parameters leads to large variations in $M_{\rm disk}$.

\item Following the same logic, we consider the sequence of binaries with $M_\text{max} = 2.05\,M_\odot$ and $q=1$ for the outlier $\Gamma = 2.640$, $q=1$. The reported disk masses are $M^\text{NR}_\text{disk} = (0.05, \,0.05, \,0.19, \,1.6, \,\textbf{12.0}) \cdot 10^{-2} M_\odot$. The disk mass of our outlier increases sharply compared to the other binaries in that sequence. The lack of simulations at higher $P_{14.7}$ prevents us from reaching the same conclusion as for the other two cases, but it is quite likely that we are here close to a sharp change in $M_{\rm disk}$. We also note that for all three cases, the CEA  and REA models are as unable to capture the numerical results as our new model is.

\end{itemize}

Assuming that the sharp transitions between disk masses of $\sim 0.1M_\odot$ and $\sim 0.01M_\odot$ found in numerical simulations are indeed physical (which is quite likely if they are due in part to the collapse of the remnant to a black hole), we can now understand better the outliers in our fitting formula: they are probably in regions of parameter space where the existing fitting formulae smooth over sharp changes in $M_{\rm disk}$ as a function of the input parameters, but where the sparsity of numerical results prevent us from reliably developing a better fit. 

From Fig.~\ref{fig:nr_vs_fit}, we can see that despite its simplicity, our formula compares well to the results of REA and CEA~\cite{2018ApJ...869..130R,Coughlin:2018miv}, although of course that comparison is biased by the fact that $M_{\rm disk}^{\rm REA}$ and $M_{\rm disk}^{\rm CEA}$ are only fitted to one part of the numerical dataset used in our study. There is no particular improvement over the previously published disk mass formulae for the binary systems that these models are calibrated on. More importantly, we expect that, due to this enlarged dataset, our formula will perform well for a broader range of parameters, in particular a wider range of mass ratios and a larger variety of total binary masses.

\section{Dynamical Ejecta for BNS Binaries} \label{sec:nsns_ejectamass}

We now move to predictions for the amount of mass ejected by BNS binaries within a few milliseconds of the merger, or {\it dynamical ejecta}. Dietrich \& Ujevic~\cite{2017CQGra..34j5014D} gathered 172 numerical simulations of BNS binaries to construct what remains the most accurate estimate of the dynamical ejecta produced in BNS merger simulations.\footnote{These simulations are distinct from the ones used in the previous section to fit the mass of the remnant disk. 
Dietrich \& Ujevic~\cite{2017CQGra..34j5014D} does not report remnant disk masses (a quantity that can be hard to define for many simulations), and does not use the results of~\cite{2017CQGra..34j5014D,Kiuchi:2019lls} as it precedes the publication of these simulations.}
They find that the mass of dynamical ejecta, $M_{\rm dyn}^{\rm DU}$, is
\begin{eqnarray}
\frac{M^{\rm DU}_{\rm dyn}}{10^{-3}M_\odot}
&=&
\left[a\left(\frac{M_1}{M_2}\right)^{1/3} \frac{1-2C_1}{C_1}+b\left(\frac{M_2}{M_1}\right)^n+ \right. \nonumber \\
&& \left. c \left(1-\frac{M_1}{M_1^b} \right)\right] M_1^b + (1 \xleftrightarrow{} 2) + d
\end{eqnarray}
with $M_1^b$ the baryon mass of neutron star $1$, $C_1$ its compactness, and $M_1$ its gravitational mass. Negative values are interpreted as $M_{\rm dyn}=0$. The free coefficients, fitted to numerical simulations, are $a=-1.35695, b=6.11252, c=-49.4355, d=16.1144$, and $n=-2.5484$. Variations of this formula fitted to the same numerical data but removing the dependence in $M_1^b$ and calculating errors in $\log{(M_{\rm dyn})}$ instead of $M_{\rm dyn}$ have been used instead in~\cite{2019MNRAS.489L..91C,Coughlin:2018miv}. The error in these fitting formulae are quite large ($\sim (0.005-0.01)M_\odot$, which is comparable to the amount of matter ejected). Whether this is due to unmodeled physical effects or finite-resolution errors in numerical simulations remains uncertain. 

The functional form of this formula is strongly inspired from previous work on black hole-neutron star binaries~\cite{Foucart:2012nc,Kawaguchi:2016ana}, where the first term is proportional to the estimated disruption radius of the neutron star and the overall functional form is motivated by the physics of tidal disruption events. However, in BHNS binaries, the coefficient $a>0$, indicating that a neutron star disrupting at a large distance from its companion favors mass ejection (and disk formation), while here the best fit results imply $a<0$. This provides an acceptable fit to the numerical results, but takes away the most natural physical interpretation of that term and may lead to more issues when {\it extrapolating} results outside of the range of existing numerical simulations. While extrapolation of fitting formulae is always a dangerous exercise, it is sometimes necessary when these formulae are used to make predictions over the entire parameter space compatible with an observed event. 

This is mainly an issue for the formulae used in~\cite{2019MNRAS.489L..91C,Coughlin:2018miv}, which still have $a<0, b>0$ but effectively set $c=0$. This implies $dM_{\rm dyn}/dC_1>0, dM_{\rm dyn}/dC_2>0$ for all $C_1,C_2$. On the other hand, we know that very compact stars promptly collapse to a black hole at merger, and have $M_{\rm dyn}=0$. So while these fitting formulae perform well within the narrow range of parameters where numerical relativity simulations are available, they also have an erroneous behavior for compact stars: they predict that the most compact stars eject the most material. The original formula from Dietrich \& Ujevic does not suffer from this issue as strongly because $M_1/M_1^b<1$ becomes smaller for more compact stars, and $c$ is large and negative.

\begin{figure}[htbp]
    \centering
    \includegraphics[width=0.45\textwidth]{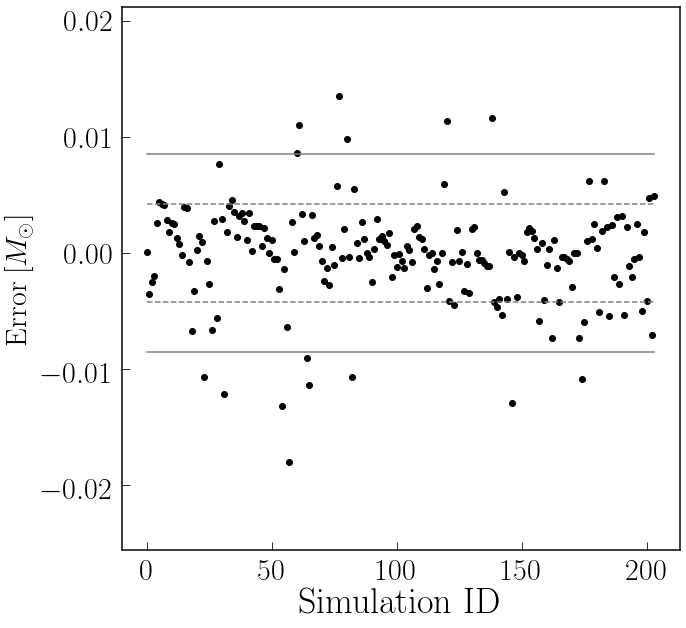}
    \caption{Difference between our fitting formula for the dynamical ejecta of binary neutron star mergers and numerical data. The first 172 simulations (black dots) are from Table I of~\cite{2017CQGra..34j5014D}; the last 28 simulations (red crosses) are from Table I of~\cite{Kiuchi:2019lls}. Dashed and solid vertical lines are $1-\sigma$ and $2-\sigma$ ranges of a zero-mean Gaussian fitted to the error distribution. Our results are very similar to Fig.2 of~\cite{2017CQGra..34j5014D}, with the addition of the more asymmetric simulations from~\cite{Kiuchi:2019lls}.}
     \label{fig:BNSFitError}
\end{figure}

We can however construct an estimate for $M_{\rm dyn}$ that is as accurate as Dietrich \& Ujevic within the range of binary parameters covered by existing numerical simulations, and relies on a simpler functional form that does not require knowledge of the baryon mass of the neutron stars. We assume
\begin{eqnarray}
    \frac{M_{\rm dyn}}{10^{-3}M_\odot} = \left(\frac{a}{C_1} + b \frac{M_2^n}{M_1^n} + c C_1\right) M_1 + (1 \xleftrightarrow{} 2)
    \label{eq:nsns_ejecta}
\end{eqnarray}
and find best-fit coefficients $a=-9.3335, b=114.17, c=-337.56$, and $n=1.5465$.\footnote{Note that the coefficients presented here are a fit to both the data from~\cite{2017CQGra..34j5014D} and additional recent results from Kiuchi \etal\cite{Kiuchi:2019lls}, to take advantage of the additional exploration of neutron star merger close to the threshold mass for collapse to a black hole performed in~\cite{Kiuchi:2019lls}. The quality of the fit does not change if we limit ourselves to the result of~\cite{2017CQGra..34j5014D}, but the best-fit coefficients vary at the $2\%$ level, depending on which data is taken into account.} 
As before, negative values imply $M_{\rm dyn}=0$.
This formula predicts a maximum in $M_{\rm dyn}(C_1)$, $M_{\rm dyn}(C_2)$ at values of $C_1,C_2$ within the physical range of compactness for neutron stars, and no matter outflows for either very compact or very large stars, as observed in numerical simulations so far (large stars however do lead to the formation of more massive disks, as discussed in the previous section, and will thus eject matter at later times in the form of disk winds). Fitting a Gaussian to the residuals of the fit, we find that the numerical results have a standard deviation $\sigma=0.004M_\odot$ with respect to the fitting formula. Differences between the numerical data and the fitting formula for the 200 simulations from~\cite{2017CQGra..34j5014D,Kiuchi:2019lls} are shown on Fig.~\ref{fig:BNSFitError}. We add the 28 simulations from~\cite{Kiuchi:2019lls} to the calibration data, as~\cite{Kiuchi:2019lls} has the advantage of including both very asymmetric mergers and mergers close to the threshold mass for rapid collapse of the remnant to a black hole.

\begin{figure*}[htbp]
    \centering
    \includegraphics[width=0.9\textwidth]{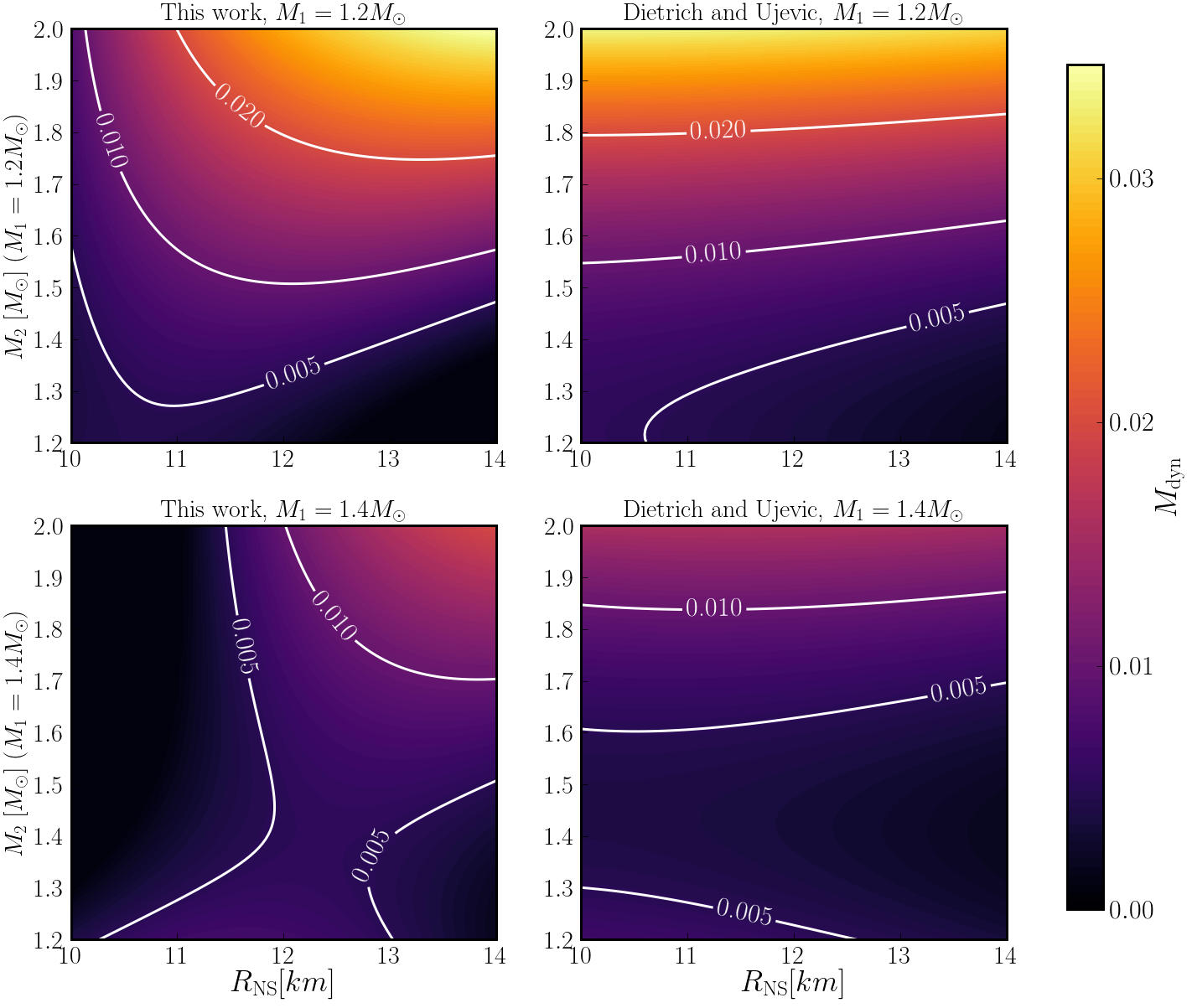}
    \caption{Mass of dynamical ejecta for binary neutron star mergers using the results from this work, as well as from~\cite{2017CQGra..34j5014D}. We assume that $M_1=1.2M_\odot$ (top) or $M_1=1.4M_\odot$ (bottom) and that both neutron stars have radius $R_{\rm NS}$. The main difference between the two fitting formulae is their behavior for compact stars, where we predict negligible mass ejection. This appears more consistent with the few available numerical simulations in that regime, and physically-motivated expectations for the rapid collapse of the post-merger remnant for very compact stars.}
     \label{fig:BNSMdyn}
\end{figure*}

Whether one uses the fitting formula from Dietrich \& Ujevic or the one presented here, the main lesson learnt is probably that all predictions have large relative uncertainties. To illustrate this, we show in Fig.~\ref{fig:BNSMdyn} the predictions from both our fitting formula and the formula from Dietrich \& Ujevic, setting for concreteness $M_1=1.2M_\odot$ (or $M_1=1.4M_\odot$), $R_1=R_2$, and following the approximation from~\cite{Lattimer:2000nx},
\begin{equation}
    M^b = M \left(1+\frac{0.6C}{1-0.5C}\right).
\end{equation}
We see that the two fitting formulae are in qualitative agreement for large neutron star radii, but have distinct behaviors for high compactness, where few numerical simulations are available. 
While we do expect high-compactness neutron stars to eject a negligible amount of matter, as predicted by our fitting formula, without calibration to numerical simulations in the correct regime it is impossible to know how accurate these predictions are. 

\section{Dynamical Ejecta for BHNS binaries} \label{sec:bhns_ejectamass}

Some of the issues that we have just discussed regarding analytical formulae predicting the mass of dynamical ejecta in neutron star binaries are also worth studying in the case of black hole-neutron star binaries. For mixed binaries, the best existing predictions for the mass of material ejected at the time of merger can be found in Kawaguchi \etal\cite{Kawaguchi:2016ana}. In that work, the mass of the dynamical ejecta, $M_{\rm dyn}^{\rm KKST}$, is modeled using the functional form
\begin{align}
\frac{M_{\rm dyn}^{\rm KKST}}{M_{\rm NS}^b}
& = a_1 Q^{n_1} \frac{1-2C_{\rm NS}}{C_{\rm NS}} - a_2 Q^{n_2} \frac{R_{\rm ISCO}}{M_{\rm BH}} \nonumber\\
& \quad +a_3 \left(1-\frac{M_{\rm NS}}{M_{\rm NS}^b}\right)+a_4,
\end{align}
with $Q=M_{\rm BH}/M_{\rm NS}$ the mass ratio of the binary, $C_{\rm NS}=GM_{\rm NS}/(R_{\rm NS}c^2)$, and $R_{\rm ISCO}$ the radius of the innermost stable circular orbit for test particles around a black hole of mass $M_{\rm BH}$ and spin equal to the component of the black hole spin {\it aligned with the orbital angular momentum}. As usual, negative values should be interpreted as $M_{\rm dyn}=0$. Fitting to 45 numerical simulations in~\cite{Kawaguchi:2016ana} led to the choice of coefficients $a_1=0.04464, a_2=0.002269, a_3=2.431, a_4=-0.4159, n_1=0.2497$, and $n_2=1.352$. This formula is accurate to $\sim 20\%$ (or $\sim 0.01M_\odot$ for low $M_{\rm dyn}$) within the range of numerical simulations used for the fit ($Q\sim 3-7$, aligned component of the dimensionless black hole spin $\chi_{\rm eff}\sim 0-0.75$, $C_{\rm NS}\sim 0.14-0.18$), including for precessing binaries~\cite{Kawaguchi:2015bwa}, simulations independently performed with a different code~\cite{Foucart:2016vxd}, and even when extrapolated to $Q\sim 1$~\cite{Foucart:2019bxj}. It has thus been remarkably successful at predicting dynamical mass ejection from black hole-neutron star binaries.\footnote{The mass of dynamical ejecta in black hole-neutron star mergers is typically higher than for binary neutron star mergers, at least when the neutron star disrupts, and thus $0.01M_\odot$ of uncertainty in the mass of dynamical ejecta is a much more satisfactory result for mixed binaries than for double neutron star systems.}

\begin{figure}[htbp]
    \centering
    \includegraphics[width=0.45\textwidth]{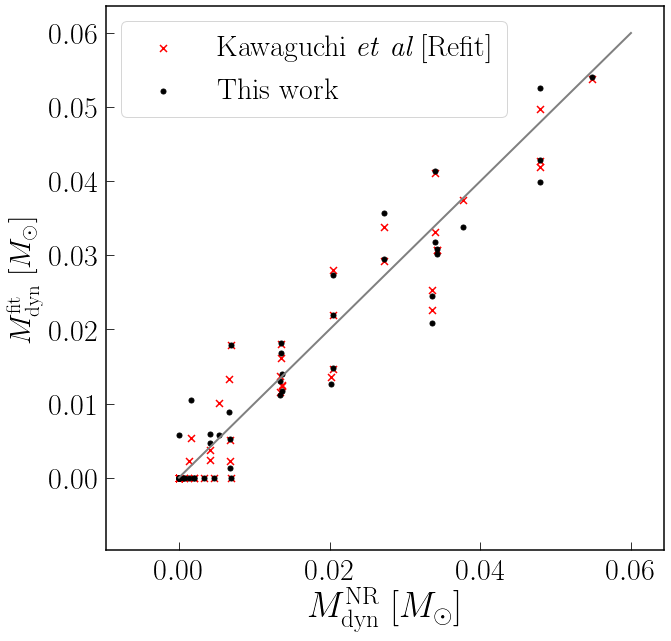}
    \caption{Fitting formulae for the dynamical ejecta of black hole-neutron star mergers plotted against numerical results for the same binary parameters. We show results for a refit of the formula from~\cite{Kawaguchi:2015bwa} and our new results. The numerical data is from~\cite{Kawaguchi:2015bwa,Foucart:2019bxj}.}
     \label{fig:BhNsFitError}
\end{figure}

Despite its success, this formula does have an important drawback when used as a black box to interpret joint gravitational wave and electromagnetic observations of black hole-neutron star binaries: its behavior for compact stars. At constant $(Q,\chi_{\rm eff})$, the formula predicts that $M_{\rm dyn}$ has a minimum value at a given compactness, and increases with both decreasing and increasing neutron star radius. This can lead to unphysical predictions: for example, a kilonova observation requiring a significant amount of ejected material could be deemed compatible with an equation of state producing very compact stars, even though physically those stars do not disrupt.
As for binary neutron star systems, we thus propose an alternative fitting formula that has the correct physical behavior for neutron stars of high compactness / small radius. Noting that the third term in the original formula is responsible for the rise of $M_{\rm dyn}$ for compact star, we take the ansatz
\begin{align}
    \frac{M_{\rm dyn}}{M_{\rm NS}^b}
        & = a_1 Q^{n_1} \frac{1-2C_{\rm NS}}{C_{\rm NS}} - a_2 Q^{n_2} \frac{R_{\rm ISCO}}{M_{\rm BH}} + a_4,
    \label{eq:bhns_ejecta}
\end{align}
which has both the correct asymptotic behavior and less free coefficients. Fitting to the simulations results from~\cite{Kawaguchi:2015bwa,Foucart:2019bxj}, we get $a_1=0.007116, a_2=0.001436, a_4=-0.02762, n_1=0.8636$, and $n_2=1.6840$. Defining the numerical error as
\begin{equation}
\Delta M_{\rm dyn}^{\rm NR} = \sqrt{(0.1M_{\rm dyn}^{\rm NR})^2 + (0.01M_\odot)^2 }    
\end{equation}
the best fit coefficients have a reduced $\chi^2_r=0.22$. Refitting the ansatz from Kawaguchi \etal instead would lead to a slightly better quality of fit, $\chi^2_r\sim 0.19$, but worse behavior outside of the fitting region. The Kawaguchi \etal formula is also a slightly better fit if we proceed as for binary neutron star mergers and fit a zero-mean Gaussian to the residuals of the fit: we find $\sigma=0.0042M_\odot$ if we refit the ansatz from Kawaguchi \etal to our full dataset, and $\sigma=0.0047M_\odot$ with our new ansatz. A visualization of fitting errors for our full dataset is provided on Fig.~\ref{fig:BhNsFitError}.

\begin{figure*}[htbp]
    \centering
    \includegraphics[width=0.9\textwidth]{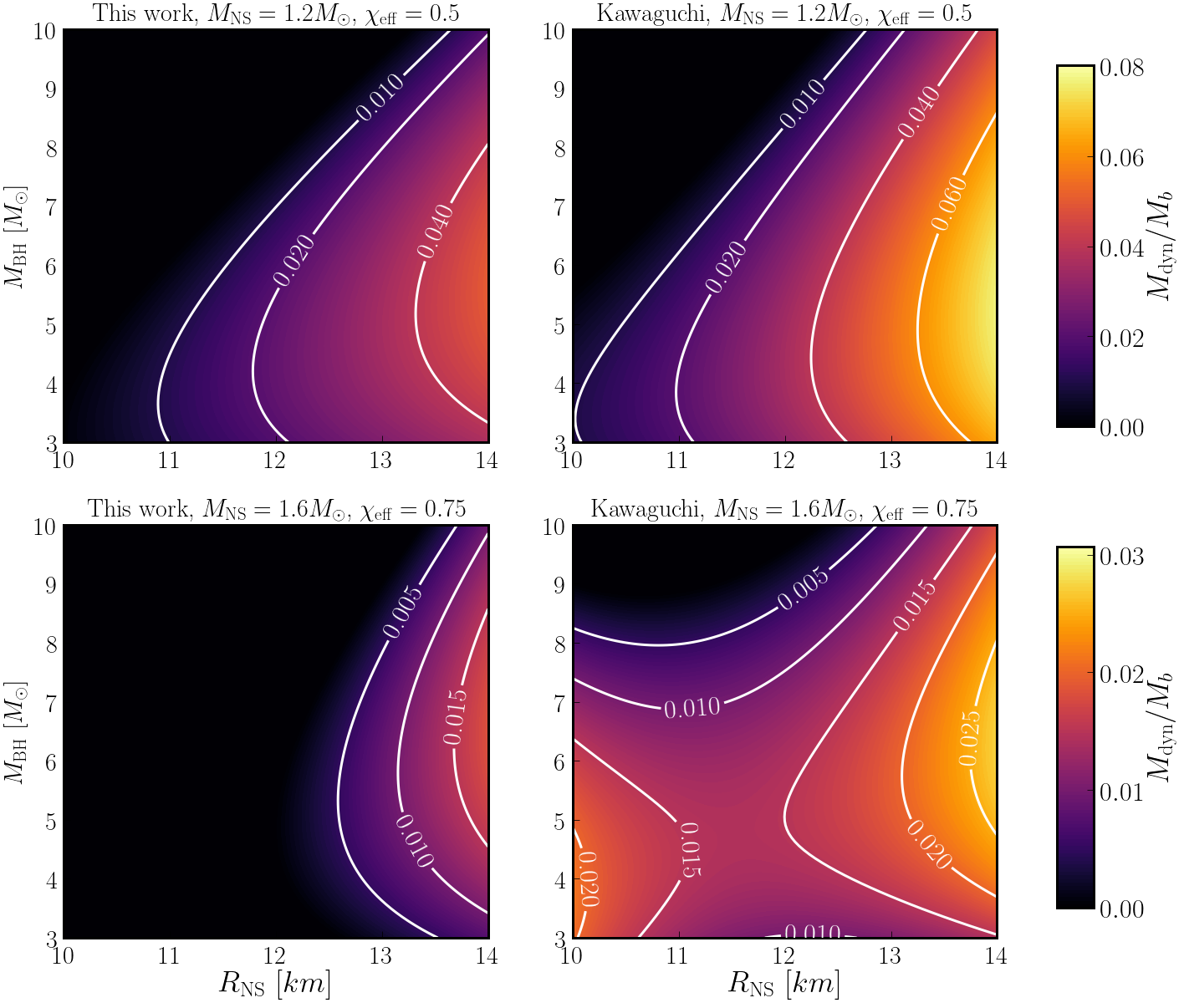}
    \caption{Mass of dynamical ejecta for black hole-neutron star mergers using the results from this work, as well as from~\cite{Kawaguchi:2015bwa}. We assume that $M_1=1.2M_\odot$, $\chi_{\rm eff}=0.5$ (top) or $M_1=1.6M_\odot$, $\chi_{\rm eff}=0.75$ (bottom). The first case is a regime well-tested in numerical simulations, where~\cite{Kawaguchi:2015bwa} performs well, while the second shows that fitting formula's issues for compact neutron stars.}
     \label{fig:BHNSMdyn}
\end{figure*}

The difference between the two fitting formulae is illustrated by Fig.~\ref{fig:BHNSMdyn}, for a region where both provide similar physical results ($M_{\rm NS}=1.2M_\odot$, $\chi_{\rm eff}=0.5$), and the results from~\cite{Kawaguchi:2015bwa} are likely to be slightly more accurate; and a region where the non-monotonic behavior of $M_{\rm dyn}$ as a function of $C_{\rm NS}$ in~\cite{Kawaguchi:2015bwa} becomes problematic ($M_{\rm NS}=1.6M_\odot$, $\chi_{\rm eff}=0.75$). The left side of that last figure corresponds to non-disrupting neutron stars, where we expect $M_{\rm dyn}=0$. We emphasize again that this only happens because we are using~\cite{Kawaguchi:2015bwa} outside of its nominal region of validity -- the original formula works perfectly well for neutron stars with radii within the range used by numerical simulations so far. Our updated formula is slightly less accurate in regions covered by numerical simulations, but has the advantage of providing accurate predictions in regions of parameter space where we do not have numerical data, but know what the correct answer should be ($M_{\rm dyn}=0$).

\section{Conclusions} \label{sec:conclusions}

We consider predictions for the disk mass of BNS binaries as well as the mass of the dynamical ejecta from BNS and BHNS binaries based on the results of numerical simulations. For all three cases we discussed the accuracy and limitations of established fitting formulae presented in published literature; not too surprisingly, we find that existing formulae work well in the region of parameter space where they were calibrated to simulations, while providing occasionally problematic predictions outside of their nominal region of validity. Similar limitations of our own formula for the mass remaining outside of the black hole after a BHNS merger had already led us recently to revise that fit~\cite{Foucart:2012nc,2018PhRvD..98h1501F}. Here, taking advantage of new numerical simulations and of some physical considerations for the outcome of the merger of very compact stars, we propose new fitting formulae for the three above mentioned cases, Eqs.~\eqref{eq:nsns_disk}, \eqref{eq:nsns_ejecta}, and \eqref{eq:bhns_ejecta}, which not only possess an analytically simpler structure than already existing fitting formulae but also provide realistic estimates for the disk mass or ejecta mass over a significantly larger portion of the parameter space.

For the disk mass of BNS binaries, two previously published fitting formulae (cf. REA and CEA) relied on the strong correlation of the disk mass with the binary tidal deformability or with the threshold mass of the binary system for equal mass systems. Our proposed formula, Eq.~\eqref{eq:nsns_disk}, relies on another physically reasonable correlation, that is more readily apparent for unequal mass binaries, between the remnant disk mass and the compactness of the lighter of the two neutron stars. Our proposed formula is simple and effective for astrophysically relevant scenarios, i.e. when the neutron star compactness exceeds the lower bound of $C \gtrsim 0.12$, including in the equal mass regime. It generally reproduces results from numerical simulations to an accuracy of better than $35\,\%$ for binaries with a broad range of mass ratios (the formula is calibrated using simulations with mass ratios as low as $q = 0.775$), total masses and binary tidal deformabilities. We note however that even our new formula fails to capture the outcome of $3$ numerical simulations in a region of parameter space where rapid changes in $M_{\rm disk}$ cannot be reliably modeled without, most likely, a denser grid of numerical simulations. An important difference between our formula and pre-existing results is its behavior for massive, unequal mass systems: our formula is more favorable to the formation of massive accretion disks in such systems. This result is partially supported by recent simulation results presented in Kiuchi \etal\cite{Kiuchi:2019lls}, and partially by our expectation that a high mass ratio BNS system with a very compact neutron star as its most massive component would not behave very differently from a disrupting BHNS system at the same mass ratio. Results for high-mass, asymmetric systems could be particularly important when assessing the potential for EM signals from systems such as GW190425~\cite{2020arXiv200104474K,2020MNRAS.tmp..680F}. However, we should caution that none of the numerical simulations used to calibrate our formula have both unequal component masses and a total mass of more than $3M_\odot$. Accordingly, using this formula (or any of the already existing predictions) to predict the outcome of GW190425 requires extrapolation of the formula into a yet-untested regime. While comparison to BHNS results provide some justification for our formula in the regime of high-mass and very asymmetric systems, there is no particular reason for it to perform better for high-mass, symmetric systems---and no way to determine where the boundary between these two regions lies without more numerical simulations in that poorly explored region of parameter space.

Our proposed formula for the ejecta mass, $M_{\rm dyn}$, of BNS binaries yields an accuracy comparable to the existing formula from Dietrich \& Ujevic~\cite{2017CQGra..34j5014D}. However, it has a simpler functional form and does not require the knowledge of the baryon mass of the neutron stars. The fitting formula also accounts for the expectation that binaries with very compact or very large neutron stars produce only negligible amounts of dynamical ejecta. The calibration data for this fitting formula are taken from 200 binary numerical simulations---including a number of simulations from binaries with a strong mass asymmetry or which are close to the threshold mass for rapid collapse.

Finally, for the dynamical ejecta of BHNS binaries, the existing fitting formula from Kawaguchi \etal\cite{Kawaguchi:2016ana} has been calibrated to merger simulations covering an extensive part of the parameter space and has proven successful so far. However, it comes with the drawback of predicting unphysically large amounts of ejecta from binaries with a very compact neutron star (owing to the fact that the formula was not intended to be used in that region of the parameter space). We isolate and remove the term responsible for this behaviour and propose a new fitting formula that displays a slightly less accurate fit to the existing dataset, but with the advantage of providing physically more reasonable estimates for BHNS binaries that contain a very compact neutron star.

Overall, our three fitting formulae can be seen as another iteration in the process of finding accurate, yet simple models for disk and ejecta masses in binary mergers of compact objects.

\begin{acknowledgements}
The authors are grateful to Geert Raaijmakers, Samaya Nissanke and Tanja Hinderer for useful discussions and suggestions regarding this project.
C.K. acknowledges support from the DFG reserach grant 413873357.
F.F. gratefully acknowledges support from
NASA through grant 80NSSC18K0565, from the
NSF through grant PHY1806278, and from the DOE through CAREER grant DE-SC0020435.
\end{acknowledgements}

\appendix*

\section{Collected data from BNS simulations}
\label{sec:appendix}
    
In this Appendix, we provide a comprehensive list in Table~\ref{tab:bns_diskmass_data} of the simulation data that were used to calibrate the fitting formula for the disk mass of double neutron star binaries.

\onecolumngrid

\begin{table}
    \caption{Data from NR simulations from Radice \etal\cite{2018ApJ...869..130R} (upper part) and Kiuchi \etal\cite{Kiuchi:2019lls} (lower part) which are used for the calibration of the fitting formula for the disk mass of BNS binaries. $M_1$, $M^b_1$, and $C_1$ are the gravitational mass, baryon mass, and compactness of the lighter neutron star, respectively (and accordingly for star 2); $q = M_1/M_2$ is the mass ratio of the binary; $M^{\rm NR}_{\rm disk}$ is the disk mass observed in the numerical simulation; $M_{\rm tot} = M_1+M_2$ the total mass of the binary; $M_{\rm thr}$ is the threshold mass \cite{2013PhRvL.111m1101B}; $\tilde{\Lambda}$ is the binary tidal deformability. The two neutron stars are labelled such that $M_1 \leq M_2$.}
    \begin{tabular}{lllllllld{2.2}llr}
    \toprule
    \mc{Model} & \mc{$M_1$} & \mc{$M^b_1$} & \mc{$C_1$} & \mc{$M_2$} & \mc{$M^b_2$} & \mc{$C_2$} & \mc{$q$} & \mc{$10^2 M^\text{NR}_\text{disk}$} & \mc{$M_\text{tot}$} & \mc{$M_\text{thr}$} & \mc{$\tilde{\Lambda}$} \\
    \midrule
$\texttt{BHBlp\_M1365125\_LK}$ & 1.25 & 1.351 & 0.140 & 1.365 & 1.489 & 0.153 & 0.9158 & 18.73 & 2.615 & 3.20 & 1028 \\
$\texttt{BHBlp\_M135135\_LK}$  & 1.35 & 1.471 & 0.151 & 1.35  & 1.471 & 0.151 & 1.0000 & 14.45 & 2.7 & 3.20 & 857 \\
$\texttt{BHBlp\_M140120\_LK}$  & 1.2  & 1.293 & 0.135 & 1.4   & 1.531 & 0.156 & 0.8571 & 20.74 & 2.6 & 3.20 & 1068 \\
$\texttt{BHBlp\_M140140\_LK}$  & 1.4  & 1.531 & 0.156 & 1.4   & 1.531 & 0.156 & 1.0000 & 7.05 & 2.8 & 3.20 & 697 \\
$\texttt{BHBlp\_M144139\_LK}$  & 1.39 & 1.519 & 0.155 & 1.44  & 1.580 & 0.161 & 0.9653 & 8.28 & 2.83 & 3.20 & 655 \\
$\texttt{BHBlp\_M150150\_LK}$  & 1.5  & 1.653 & 0.167 & 1.5   & 1.653 & 0.167 & 1.0000 & 1.93 & 3 & 3.20 & 462 \\
$\texttt{BHBlp\_M160160\_LK}$  & 1.6  & 1.777 & 0.178 & 1.6   & 1.777 & 0.178 & 1.0000 & 0.09 & 3.2 & 3.20 & 306 \\
$\texttt{DD2\_M1365125\_LK}$   & 1.25 & 1.351 & 0.140 & 1.365 & 1.489 & 0.153 & 0.9158 & 20.83 & 2.615 & 3.35 & 1028 \\
$\texttt{DD2\_M135135\_LK}$    & 1.35 & 1.471 & 0.151 & 1.35  & 1.471 & 0.151 & 1.0000 & 15.69 & 2.7 & 3.35 & 858 \\
$\texttt{DD2\_M140120\_LK}$    & 1.2  & 1.293 & 0.135 & 1.4   & 1.531 & 0.156 & 0.8571 & 19.26 & 2.6 & 3.35 & 1070 \\
$\texttt{DD2\_M140140\_LK}$    & 1.4  & 1.531 & 0.156 & 1.4   & 1.531 & 0.156 & 1.0000 & 12.36 & 2.8 & 3.35 & 699 \\
$\texttt{DD2\_M144139\_LK}$    & 1.39 & 1.519 & 0.155 & 1.44  & 1.580 & 0.161 & 0.9653 & 14.40 & 2.83 & 3.35 & 658 \\
$\texttt{DD2\_M150150\_LK}$    & 1.5  & 1.653 & 0.167 & 1.5   & 1.653 & 0.167 & 1.0000 & 16.70 & 3 & 3.35 & 469 \\
$\texttt{DD2\_M160160\_LK}$    & 1.6  & 1.777 & 0.178 & 1.6   & 1.777 & 0.178 & 1.0000 & 1.96 & 3.2 & 3.35 & 317 \\
$\texttt{LS220\_M120120\_LK}$  & 1.2  & 1.309 & 0.138 & 1.2   & 1.309 & 0.138 & 1.0000 & 17.43 & 2.4 & 3.05 & 1439 \\
$\texttt{LS220\_M1365125\_LK}$ & 1.25 & 1.369 & 0.144 & 1.365 & 1.508 & 0.158 & 0.9158 & 16.86 & 2.615 & 3.05 & 848 \\
$\texttt{LS220\_M135135\_LK}$  & 1.35 & 1.490 & 0.157 & 1.35  & 1.490 & 0.157 & 1.0000 & 7.25 & 2.7 & 3.05 & 684 \\
$\texttt{LS220\_M140120\_LK}$  & 1.2  & 1.309 & 0.138 & 1.4   & 1.551 & 0.163 & 0.8571 & 22.82 & 2.6 & 3.05 & 893 \\
$\texttt{LS220\_M140140\_LK}$  & 1.4  & 1.551 & 0.163 & 1.4   & 1.551 & 0.163 & 1.0000 & 4.58 & 2.8 & 3.05 & 536 \\
$\texttt{LS220\_M144139\_LK}$  & 1.39 & 1.539 & 0.162 & 1.44  & 1.600 & 0.168 & 0.9653 & 3.91 & 2.83 & 3.05 & 499 \\
$\texttt{LS220\_M145145\_LK}$  & 1.45 & 1.613 & 0.169 & 1.45  & 1.613 & 0.169 & 1.0000 & 2.05 & 2.9 & 3.05 & 421 \\
$\texttt{LS220\_M150150\_LK}$  & 1.5  & 1.675 & 0.176 & 1.5   & 1.675 & 0.176 & 1.0000 & 0.16 & 3 & 3.05 & 331 \\
$\texttt{LS220\_M160160\_LK}$  & 1.6  & 1.801 & 0.189 & 1.6   & 1.801 & 0.189 & 1.0000 & 0.07 & 3.2 & 3.05 & 202 \\
$\texttt{LS220\_M171171\_LK}$  & 1.71 & 1.944 & 0.205 & 1.71  & 1.944 & 0.205 & 1.0000 & 0.06 & 3.42 & 3.05 & 116 \\
$\texttt{SFHo\_M1365125\_LK}$  & 1.25 & 1.363 & 0.154 & 1.365 & 1.503 & 0.169 & 0.9158 & 8.81 & 2.615 & 2.95 & 520 \\
$\texttt{SFHo\_M135135\_LK}$   & 1.35 & 1.485 & 0.167 & 1.35  & 1.485 & 0.167 & 1.0000 & 6.23 & 2.7 & 2.95 & 422 \\
$\texttt{SFHo\_M140120\_LK}$   & 1.2  & 1.302 & 0.148 & 1.4   & 1.546 & 0.174 & 0.8571 & 11.73 & 2.6 & 2.95 & 546 \\
$\texttt{SFHo\_M140140\_LK}$   & 1.4  & 1.546 & 0.174 & 1.4   & 1.546 & 0.174 & 1.0000 & 0.01 & 2.8 & 2.95 & 334 \\
$\texttt{SFHo\_M144139\_LK}$   & 1.39 & 1.533 & 0.172 & 1.44  & 1.596 & 0.179 & 0.9653 & 0.09 & 2.83 & 2.95 & 312 \\
$\texttt{SFHo\_M146146\_LK}$   & 1.46 & 1.621 & 0.182 & 1.46  & 1.621 & 0.182 & 1.0000 & 0.02 & 2.92 & 2.95 & 252 \\
\midrule
$\Gamma = 3.765$, $q = 1.0$ & 1.375 & 1.551 & 0.195 & 1.375 & 1.551 & 0.195 & 1.000 & 0.05 & 2.75 & 2.72 & 208 \\
$\Gamma = 3.765$, $q = 0.775$ & 1.2 & 1.331 & 0.172 & 1.55 & 1.779 & 0.219 & 0.775 & 2.3 & 2.75 & 2.72 & 218 \\
$\Gamma = 3.887$, $q = 1.0$ & 1.375 & 1.550 & 0.194 & 1.375 & 1.550 & 0.194 & 1.000 & 0.05 & 2.75 & 2.76 & 221 \\
$\Gamma = 3.887$, $q = 0.775$ & 1.2 & 1.331 & 0.171 & 1.55 & 1.778 & 0.171 & 0.775 & 2.9 & 2.75 & 2.76 & 230 \\
$\Gamma = 4.007$, $q = 1.0$ & 1.375 & 1.550 & 0.193 & 1.375 & 1.550 & 0.193 & 1.000 & 0.27 & 2.75 & 2.79 & 232 \\
$\Gamma = 3.446$, $q = 1.0$ & 1.375 & 1.544 & 0.191 & 1.375 & 1.544 & 0.191 & 1.000 & 0.05 & 2.75 & 2.76 & 232 \\
$\Gamma = 3.446$, $q = 0.775$ & 1.2 & 1.325 & 0.168 & 1.55 & 1.771 & 0.215 & 0.775 & 3.6 & 2.75 & 2.76 & 245 \\
$\Gamma = 3.568$, $q = 1.0$ & 1.375 & 1.543 & 0.190 & 1.375 & 1.543 & 0.190 & 1.000 & 0.05 & 2.75 & 2.80 & 247 \\
$\Gamma = 3.568$, $q = 0.775$ & 1.2 & 1.325 & 0.167 & 1.55 & 1.770 & 0.213 & 0.775 & 3.8 & 2.75 & 2.80 & 259 \\
$\Gamma = 3.687$, $q = 1.0$ & 1.375 & 1.543 & 0.189 & 1.375 & 1.543 & 0.189 & 1.000 & 0.78 & 2.75 & 2.83 & 260 \\
$\Gamma = 3.132$, $q = 1.0$ & 1.375 & 1.534 & 0.185 & 1.375 & 1.534 & 0.185 & 1.000 & 0.05 & 2.75 & 2.81 & 272 \\
$\Gamma = 3.132$, $q = 0.775$ & 1.2 & 1.318 & 0.161 & 1.55 & 1.759 & 0.209 & 0.775 & 6.3 & 2.75 & 2.81 & 290 \\
$\Gamma = 3.252$, $q = 1.0$ & 1.375 & 1.535 & 0.184 & 1.375 & 1.535 & 0.184 & 1.000 & 0.19 & 2.75 & 2.85 & 288 \\
$\Gamma = 3.252$, $q = 0.775$ & 1.2 & 1.319 & 0.161 & 1.55 & 1.759 & 0.207 & 0.775 & 12.0 & 2.75 & 2.85 & 305 \\
$\Gamma = 3.370$, $q = 1.0$ & 1.375 & 1.535 & 0.183 & 1.375 & 1.535 & 0.183 & 1.000 & 3.1 & 2.75 & 2.89 & 303 \\
$\Gamma = 2.825$, $q = 1.0$ & 1.375 & 1.522 & 0.176 & 1.375 & 1.522 & 0.176 & 1.000 & 1.8 & 2.75 & 2.89 & 345 \\
$\Gamma = 2.825$, $q = 0.775$ & 1.2 & 1.309 & 0.153 & 1.55 & 1.744 & 0.200 & 0.775 & 8.7 & 2.75 & 2.89 & 373 \\
$\Gamma = 2.942$, $q = 1.0$ & 1.375 & 1.523 & 0.176 & 1.375 & 1.523 & 0.176 & 1.000 & 1.6 & 2.75 & 2.93 & 362 \\
$\Gamma = 2.942$, $q = 0.775$ & 1.2 & 1.310 & 0.153 & 1.55 & 1.745 & 0.199 & 0.775 & 12.0 & 2.75 & 2.93 & 387 \\
$\Gamma = 2.528$, $q = 1.0$ & 1.375 & 1.505 & 0.163 & 1.375 & 1.505 & 0.163 & 1.000 & 5.3 & 2.75 & 3.00 & 508 \\
$\Gamma = 2.528$, $q = 0.775$ & 1.2 & 1.296 & 0.140 & 1.55 & 1.722 & 0.188 & 0.775 & 16.0 & 2.75 & 3.00 & 558 \\
$\Gamma = 2.640$, $q = 1.0$ & 1.375 & 1.508 & 0.164 & 1.375 & 1.508 & 0.164 & 1.000 & 12.0 & 2.75 & 3.63 & 516 \\
\bottomrule
\end{tabular}
    \label{tab:bns_diskmass_data}
\end{table}

\twocolumngrid

\bibliography{references}

\end{document}

%% file: preamble.tex
\usepackage[english]{babel}
\usepackage[utf8]{inputenc}
\usepackage[colorinlistoftodos,color=green!40,prependcaption]{todonotes}
\usepackage{amsthm}
\usepackage{mathtools}
\usepackage{physics}
\usepackage{xcolor}
\usepackage{graphicx}
\usepackage[left=23mm,right=13mm,top=35mm,columnsep=15pt]{geometry} 
\usepackage{adjustbox}
\usepackage{placeins}
\usepackage[T1]{fontenc}
\usepackage{lipsum}
\usepackage{csquotes}
\usepackage{booktabs}
\usepackage{tabularx}
\usepackage{ltablex}
\usepackage{longtable}
\usepackage{soul}
\usepackage{cancel}
\usepackage{journals}
\usepackage{enumitem} 
\usepackage{dcolumn}

\usepackage[pdftex, pdftitle={Article}, pdfauthor={Author}]{hyperref} 

\newcommand{\unit}[1]{%
    \ensuremath{\, \mathrm{#1}}}
    
\newcommand{\etal}{{\it et al. }}

\newcolumntype{d}[1]{D{.}{.}{#1}}
\newcommand{\mc}[1]{\multicolumn{1}{c}{#1}}